\documentclass[twocolumn,aps,prl,superscriptaddress,showpacs,secnumroman,showkeys]{revtex4-1}
\usepackage{amsmath,bm}%
\usepackage{graphicx}%
\usepackage{color}%

\begin{document}

% Use the \preprint command to place your local institutional report
% number in the upper righthand corner of the title page in preprint mode.
% Multiple \preprint commands are allowed.
% Use the 'preprintnumbers' class option to override journal defaults
% to display numbers if necessary
%\preprint{}

%Title of paper
\title{Effect of compound nucleus spin correlations on fission fragment angular distribution}

% repeat the \author .. \affiliation  etc. as needed
% \email, \thanks, \homepage, \altaffiliation all apply to the current
% author. Explanatory text should go in the []'s, actual e-mail
% address or url should go in the {}'s for \email and \homepage.
% Please use the appropriate macro foreach each type of information

% \affiliation command applies to all authors since the last
% \affiliation command. The \affiliation command should follow the
% other information
% \affiliation can be followed by \email, \homepage, \thanks as well.
%\author{}
%\email[]{Your e-mail address}
%\homepage[]{Your web page}
%\thanks{}
%\altaffiliation{}
%\affiliation{}
\author{O.M. Gorbachenko}
\email[E-mail at:]{gorbachenko@univ.kiev.ua}
\affiliation{Faculty of Physics, Taras Shevchenko National University, Kyiv, 03680, Ukraine}
\author{S. Kun}
\email[E-mail at:]{ksy1956@gmail.com}
\affiliation{Canberra, Australia}

%Collaboration name if desired (requires use of superscriptaddress
%option in \documentclass). \noaffiliation is required (may also be
%used with the \author command).
%\collaboration can be followed by \email, \homepage, \thanks as well.
%\collaboration{}
%\noaffiliation

%\date{\today}

\begin{abstract}
We extend a conventional description of the fusion-fission fragment angular distributions by introducing the correlation
between compound nucleus states carrying different total angular momenta.
This
correlation results in the strong anisotropy and mass-angle correlation of fission fragments
for compact saddle-point nuclear shapes for which
the conventional description predicts almost isotropic angular distributions.
The spin off-diagonal phase relaxation timescale, $\simeq 10^{-19}$ sec, obtained from analysis of
anomalous fission fragment angular distributions in  $^{12}$C+$^{236}$U,  $^{16}$O+$^{232}$Th and $^{16}$O+$^{238}$U collisions
at the sub-barrier energies is three orders of magnitude longer than the timescale of the compound nucleus thermalization.
Expression for the angle-dependent time power spectrum for quasifission is also presented.
\end{abstract}

\maketitle

A conventional approach to the fission fragment angular distributions (FFAD)
\cite{BohrA56},
\cite{HalpStr58}, \cite{Ericson60} postulates vanishing of interference between fission channels populated from the compound nucleus
(CN) states with different total spins $(J)$.
This postulate is a direct consequence of the N. Bohr's hypothesis on independence of a process of the CN decay on the way it was formed.

One of the arguments against the CN  $(J_1\neq J_2 )$-correlation has been that the evaporation spectra for a decay of heavy CN are either isotropic or show
fore-aft symmetry with usually weak anisotropy \cite{Weisskopf61}. Yet there are many data sets demonstrating
a strong fore-aft asymmetry of the evaporation spectra.  The first well known example of a strong
fore-aft asymmetry of the evaporation spectra was reported in \cite{Gugelot54} for the Pt(p,p') process.
 Examples of strong fore-aft asymmetry of the evaporation proton spectra in nucleon induced reactions
with the targets $^{197}$Au, $^{208}$Pb, $^{209}$Bi and $^{nat}$U are displayed in
\cite{Benet08}. Additional data on the  strong
fore-aft asymmetry and anomalous (stronger than $(1/\sin\theta ))$ back-angle peaking of the evaporation $\alpha$-particle spectra  will be displayed
in the extended version of this Letter.

Following the arguments \cite{Weisskopf61} we associate the evaporation spectra with the CN decay and not with direct interactions. Then
the fore-aft asymmetry of the evaporation spectra provides a clear indication
of the correlation between fluctuating CN $S$-matrix elements carrying different spin values.
If so then it is not certain that the $(J_1\neq J_2 )$-correlations must
necessarily vanish for the fission of the thermalized CN. Considering fission as a complex particle
evaporation \cite{Ericson60}, \cite{Rossner84} the resulting expression \cite{Kun94},\cite{Kun97a} for the CN $(J_1\neq J_2 )$-correlation is
directly applicable to calculate FFAD.
Some aspects of the validity of the scission-point statistical
model \cite{Rossner84} have been addressed in \cite{Bertsch19a}, \cite{Bertsch19b},
\cite{Bertsch19c}.

In this Letter we extend
the more popular description (the transition-state statistical model (TSSM)) of the FFAD \cite{HalpStr58} by introducing
the CN $(J_1\neq J_2 )$-correlation.
We keep in mind that the original formulation \cite{HalpStr58} is applicable providing
the CN spin is smaller than the channel spin of the fission fragments \cite{Ericson60}, \cite{Rossner84}, \cite{Bond84}.
Instead of the standard expression \cite{HalpStr58}
we write
$W(\theta)=\sum_{K}<|\sum_{Jc}F_{a,cK}^J(\theta,E)|^2>$.
Here
$F_{a,cK}^J(\theta,E)=(2J+1)S_{a,cK}^J(E)d_{0K}^J(\theta)$,
 $<...>$ stands for the energy averaging,
$d_{0K}^{J_1}(\theta)d_{0K}^{J_2}(\theta)=D_{0K}^{J_1}(\alpha,\theta,\gamma)D_{0K}^{J_2}(\alpha,\theta,\gamma)^\ast$
and the $D$-functions are the wave functions of the axially symmetric top. In $W(\theta)$, $a$ is an index for the entrance channel,
 $K$ is projection of $J$ on the symmetry axis of the nucleus at the saddle point and $c$ stands for the rest of indices of the fission channels. $K$ off-diagonal contributions in  $W(\theta)$ are eliminated by integration over azimuthal angle around the nuclear
symmetry axis (the third Euler angle $\gamma$) at the saddle point.

Neglecting $J$-dependence of the potential phase shifts in the fission channels
we take $S$-matrix in the form $S_{a,cK}^{J}(E)=-i\exp(i\varphi_{a}^J)\delta S_{a,cK}^{J}(E)$, where $\varphi_{a}^J$ is the potential phase shift in the entrance channel while  $\delta S_{a,cK}^{J}(E)$ is given by the
pole expansion \cite{Bertsch18} with the residues $\gamma_\mu^{Ja}\gamma_\mu^{J,cK}$. Here, $\gamma$'s
are the real partial width amplitudes in the entrance and exit (fission) channels at the saddle point with the $\mu$-index denoting the CN resonance level.
We assume that statistical properties of $\gamma$'s with fixed $J$-value are described by statistics of Gaussian Orthogonal Ensemble.
Note that, for $J\geq 1$,
inclusion of $K$-projections into the fission channel indices
  takes us to a regime
 $\overline{(\gamma_\mu^{J,cK})^2}^\mu/D_J\approx 1/[2\pi(2J+1)]\ll 1$ ($D_J$ is the CN average level spacing). Therefore, the pole expansion \cite{Bertsch18}
  is an accurate approximation to the
$S$-matrix unitary representation in a regime of the strongly overlapping resonances \cite{GorinSel02}, $\Gamma /D_J\gg 1$,  with $\Gamma$
being the CN total decay width.
The $S$-matrix spin off-diagonal correlations result from the correlation between
$\gamma_{\mu_1}^{J_1 a}\sum_{c_1}\gamma_{\mu_1}^{J_1 ,c_1 K}$ and $\gamma_{\mu_2}^{J_2 a}\sum_{c_2}\gamma_{\mu_2}^{J_2 ,c_2 K}$ with $|J_1-J_2|\beta$ being
 the correlation length in the $(E_{\mu_1}^{J_1}- E_{\mu_2}^{J_2})$-space, where $E_\mu^J$ are
the resonance energies.
In turn, this correlation
originates from the correlations between the CN resonance eigenstates $\phi_{\mu}^{J}$ in the form
\begin{eqnarray}
&&{ \int\int d{\bf R_1}d{\bf R_2} }
{ \left[ \overline{\phi_{\mu_1}^{J_1}({\bf r},{\bf R_1})\phi_{\mu_1}^{J_1}({\bf r},{\bf R_2}) \phi_{\mu_2}^{J_2}({\bf r},{\bf R_1})\phi_{\mu_2}^{J_2}({\bf r},{\bf R_2})}^{\bf r} \right. } \nonumber \\
&&{\left.- \overline{\phi_{\mu_1}^{J_1}({\bf r},{\bf R_1})\phi_{\mu_1}^{J_1}({\bf r},{\bf R_2})}^{\bf r} {~}\overline{\phi_{\mu_2}^{J_2}({\bf r},{\bf R_1})\phi_{\mu_2}^{J_2}({\bf r},{\bf R_2})}^{\bf r} \right].}
\label{WaveFunCorr}
\end{eqnarray}
Here ${\bf r}=({\bf r}_1,{\bf r}_2,...,{\bf r}_n)$ are coordinates of $n$
nucleons and  ${\bf R_{1,2}}$ are
 coordinates of the rest of the $(A-n)$ nucleons.
In Eq.~\eqref{WaveFunCorr} $\overline{(...)}^{\bf r}$ stand for the spacial averaging.
Note that upon the partial $(\mu_{1},\mu_{2})$-averaging (keeping $(E_{\mu_1}^{J_1}-E_{\mu_2}^{J_2})$  fixed with accuracy $\ll \beta$)
the main contribution to Eq.~\eqref{WaveFunCorr} is produced with
${\bf R_1}\to{\bf R_2}$ driving Eq.~\eqref{WaveFunCorr} towards the correlation between
 $(\phi_{\mu_1}^{J_{1}})^2$ and $(\phi_{\mu_{2}}^{J_{2}})^2$. The widths $|J_1-J_2|\beta $ relate \cite{KunInProgress}
to the quantum analogs of imaginary parts of the Ruelle-Pollicott resonances \cite{Ruelle86}.

The final result, for the fission fragments sufficiently lighter than the average fragment mass, reads
\begin{eqnarray}
&&W(\theta)\propto\sum_{J_1,J_2} (2J_1+1)(2J_2+1) (T_a^{J_1}T_a^{J_2})^{1/2} \nonumber \\
&&\exp(i\varphi_{a}^{J_1}- i\varphi_{a}^{J_2})[1+
(\beta /\Gamma )|J_1-J_2|]^{-1} \nonumber \\
&&\sum_{K=-{\rm min}(J_1,J_2)}^{{\rm min}(J_1,J_2) } [B_{J_1}(K)B_{J_2}(K)]^{1/2}d_{0K}^{J_1}(\theta) d_{0K}^{J_2}(\theta)
\label{Wcorrect}
\end{eqnarray}
with $B_J(K >J)=0$, $B_J(K\leq J)=p_K/\sum_{{\tilde K}=-J}^{J}p_{{\tilde K}}$, $p_K=\exp[-K^2/2K_0^2]$.
Here, $K_0^2={\cal J}_{eff} T_b/\hbar^2$, ${\cal J}_{eff}^{-1}={\cal J}_{perp}^{-1}-{\cal J}_{par}^{-1}$,
${\cal J}_{par}$ and ${\cal J}_{perp}$ are the nuclear moments of inertia for rotations around the symmetry axis
and a perpendicular axis, respectively, $ T_b=[8(E-B_f-E_{rot})/A]^{1/2}$ is the nuclear temperature
at the saddle point, $B_f$ is the fission barrier, $E$ is the excitation energy and $E_{rot}$ is the
rotational energy.
In general case, the factor $[\Gamma_f^{J_1}\Gamma_f^{J_2}/(\Gamma^{J_1}\Gamma^{J_2})]^{1/2}$ is to be included in Eq.~\eqref{Wcorrect} with
$\Gamma_f^{J}$ and $\Gamma^{J}$ being $J$-dependent fission and CN total decay widths, respectively.
For $\beta /\Gamma \gg 1$ the $(J_1\neq J_2 )$-correlations decay much
faster than the CN average life-time thereby leading to the TSSM result \cite{HalpStr58}.

Eq.~\eqref{Wcorrect} generally produces fore-aft asymmetry of FFAD due to the $(J_1\neq J_2 )$-contributions with odd values of
$(J_1+J_2)$, {\sl i.e.}, $\pi_1\neq\pi_2$ ($\pi$ is parity). The fore-aft symmetry is achieved
for $\beta_{\pi_1\neq\pi_2}\gg\Gamma$ resulting in
 $W(\theta)\to W(\theta)+W(\pi-\theta)$ like for the mass-summed FFAD.
 However, such a fore-aft symmetry does not necessarily mean a complete mass(shape)-relaxation/symmetrization but implies that the
 mass-asymmetric deformed system at the saddle-point is simultaneously oriented in opposite directions with equal probabilities.

The time power spectrum, $P(\theta ,t)$, is given by Eq.~\eqref{Wcorrect} but with
$\exp[-(\Gamma +\beta |J_1-J_2|)t/\hbar]$ instead of $[1+(\beta /\Gamma )|J_1-J_2|]^{-1}$. The time-dependent FFAD are
$\propto \int_0^td\tau P(\theta ,\tau )$.

On the initial stage the colliding ions form a dinuclear system. A distribution of its orientations, $P_{K=0}(\theta )$, may be identified with the
time power spectrum at $t=0$ for heavy-ion scattering
\cite{Kun01} or for dissipative heavy-ion collisions \cite{Kun97b}, where the potential phase shifts in the exit channel being omitted.
Then we find that
$P_{K=0}(\theta )$ is formally given by Eq.~\eqref{Wcorrect} with $\beta =0$ and $K_0^2=0$ ($B_J(K)=\delta_{0K}$).
We expand
$\varphi_{a}^{J}=\Phi (J-{\bar J})+{\dot \Phi}(J-{\bar J})^2/2$, where ${\bar J}=\sum_{J=0}^\infty JT_a^J/\sum_{J=0}^\infty T_a^J $ while $\Phi$ and $\dot{\Phi}$
have a meaning of the classical deflection angle and its derivative in the entrance channel.
Then $P_{K=0}(\theta )$ has a bell shape with maximum at $\theta\simeq \Phi$ and width
$\Delta_{K=0}\simeq [(1/<J^2>)+{\dot \Phi}^2<J^2>]^{1/2}$, where $<J^2>=\sum_{J=0}^\infty (2J+1)J^2T_a^J/\sum_{J=0}^\infty (2J+1)T_a^J $
 is mean square spin value of the dinucleus. Therefore, for
$\Delta_{K=0}\leq 1$, we are dealing with a fusion of the dinuclear system preferentially orientated along the $\Phi$-direction.
It follows from Eq.~\eqref{Wcorrect} that the pre-fusion phase relations (initial conditions) in the entrance channel, $(\varphi_{a}^{J_1}- \varphi_{a}^{J_2})$, are
not forgotten for the fusion-fission even for a complete $K$-equilibration, $K_0^2\gg <J^2>$, providing  $\beta /\Gamma $ is a finite quantity.
 This means that if the same CN is formed but with different phase relations in
 the entrance channel (different collision partners) the FFAD will be different for a finite value of $\beta /\Gamma $.
For $\Delta_{K=0} \geq \pi$, the ($J_1\neq J_2$)-correlation
is strongly suppressed already on the pre-fusion stage, as can probably be the case for the above-barrier $\alpha$+$^{244}$Cm fusion-fission \cite{Vandenbosch86},\cite{Murakami86}
driving Eq.~\eqref{Wcorrect} towards the standard TSSM expression ($J_1=J_2$)
even for a finite value of $\beta /\Gamma $.

We illustrate the effect of the orientation-dependent fusion on FFAD considering the three examples.

1) $^{12}$C+$^{236}$U, $E_{c.m.}=59$ MeV, $E=34.2$ MeV \cite{Vandenbosch86}.

For $E_{c.m.}=59$ MeV, the  TSSM predicts \cite{Vandenbosch86} $A_{TSSM}=W(\pi)/W(\pi/2)= 1.1-1.15$
while the experimental value is
$A_{exp}\approx 1.7$ \cite{Vandenbosch86}, {\sl i.e.}, $(A_{exp}-1)/(A_{TSSM}-1)\approx 6$). The
4n
evaporation residues data showed no indication for a noticeable presence of the quasifission \cite{Sonzogni98}.
Moreover, to reproduce the experimental value of $\sigma_{ER}({\rm 4n})/\sigma_{fiss}\approx 3\times 10^{-3}$ for $E_{c.m.}=60$ MeV
 employing a standard statistical model it was
necessary \cite{Sonzogni98} to take $B_f=4.85$ MeV instead of the Sierk value $B_f=2.1$ MeV \cite{Sierk86}.
We have checked that, for $B_f=2.1$ MeV, a statistical model predicts  $\sigma_{ER}({\rm 4n})/\sigma_{fiss}\approx 10^{-9}$ for $E_{c.m.}=60$ MeV
instead of the experimental value of $\approx 3\times 10^{-3}$.
Since the temperature at the saddle-point is $T_b\approx 1$ MeV then, for $B_f=4.85$ MeV, the pre-equilibrium fission
\cite{Ramamurthy85},\cite{Vorkapic95},\cite{Liu96} is insignificant.

In our interpretation we take into account the orientation dependence of the sub-barrier fusion for the target having a prolate deformation \cite{Stokstad81}.
Then
we deal with a fusion of the strongly mass asymmetric dinuclear system with $K=0$ which is preferably oriented along the beam direction.
We take $\Phi=\pi$, accordingly.
Then, with $T_a^J$ obtained from Fig. 13 in \cite{Murakami86},
$P_{K=0}(\theta )$ depends on a single parameter
${\dot \Phi}$. We take $|{\dot \Phi}|=3^\circ $ resulting in a peaking of $P_{K=0}(\theta )$ at $\theta=\pi$ having a width of
$\Delta_{K=0}\approx 10^\circ$ with a tail extended up to $\approx 140^\circ $.
Note that relatively small irregular deviations from the smooth $J$-dependence of $T_a^J$ would produce long-range tail in $P_{K=0}(\theta )$.

The mass-summed FFAD is given by
Eq.~\eqref{Wcorrect} symmetrized about $\theta=\pi /2$.
In fact, no statistically significant mass-angle correlation  was observed in \cite{Murakami86}. In our language
this means  $\beta_{\pi_1\neq\pi_2}\gg\Gamma$, {\sl i.e.}, the fore-aft symmetry, $W(\theta)\to W(\theta)+W(\pi-\theta)$, independent of the fragment mass.
Since in our case $<J^2>/(4K_0^2)\approx 0.1\ll 1$ we take $K_0^2=\infty$  ($B_{J}(K)=1/(2J+1)$) resulting in the isotropic FFAD in the absence
of the $(J_1\neq J_2 )$-correlation.
For $\Phi=\pi$ and  $|{\dot \Phi}|=3^\circ $, the data fit in Fig. 1 uniquely yields $\beta/\Gamma=2.8$.
With
$\Gamma\approx \Gamma_f\approx T_b\exp(-B_f/T_b)/(2\pi )\approx 1.2$ keV ($T_b=1$ MeV, $B_f=4.85$ MeV), $\beta\approx 3.4$ keV and
$\hbar /\beta\approx 2\times 10^{-19}$ sec.
Therefore, the characteristic time for a decay of the CN  $(J_1\neq J_2 )$-correlation
is about 3 orders of magnitude longer than the thermalization time, $\hbar/\Gamma_{spr}\simeq 1.3\times 10^{-22}$ sec, where $\Gamma_{spr}\simeq 5$ MeV is a width of the giant resonances.

2) $^{16}$O+$^{232}$Th, $E_{c.m.}=74.5$ MeV, $E=38$ MeV \cite{Vandenbosch86}.

For this minimal energy measured in \cite{Vandenbosch86},\cite{Murakami86} the mass-summed $A_{exp}\approx 1.8$ (Fig. 1 in \cite{Vorkapic95}).
For $K_0^2=192$ and $T_a^J\propto \exp[-J^2/(<J^2>)]$ with $<J^2>=80$ from Fig. 12 in  \cite{Murakami86}
the TSSM ($\beta /\Gamma\to\infty$) yields
$(A_{TSSM}-1)\approx 0.1\ll (A_{exp}-1)\approx 0.8$. In order to reproduce $A_{exp}= 1.8$ with $\Phi=\pi$,  $|{\dot \Phi}|=3^\circ $ and
$K_0^2=\infty$ we have to take $\beta/\Gamma =2$, {\sl i.e.}, $\beta\approx 3$ keV.

3) $^{16}$O+$^{238}$U, $E_{c.m.}=72.8-75.6$ MeV, $E=34.5-37.3$ MeV \cite{ Hinde96}.

The  $^{16}$O+$^{238}$U evaporation residues data provided evidence \cite{Nishio04}
that the fission events are ascribed to fusion-fission and not to quasifission \cite{ Hinde95} in spite of the conclusive experimental evidence
for the mass-angle correlation \cite{ Hinde96}.
 Employing a standard statistical model we have
found that in order to reproduce the experimental value of $\sigma_{ER}({\rm 4n})/\sigma_{fiss}\approx 6\times 10^{-5}$ for $E_{c.m.}=72$ MeV it is
necessary to take $B_f=4.15$ MeV instead of the Sierk value $B_f=1.5$ MeV \cite{Sierk86}.
We have checked that, for $B_f=1.5$ MeV, a statistical model predicts  $\sigma_{ER}({\rm 4n})/\sigma_{fiss}\approx 10^{-11}$ for $E_{c.m.}=72$ MeV
instead of the experimental value of $\approx 6\times 10^{-5}$.
Since the temperature at the saddle-point is $T_b\approx 1$ MeV then, for
$B_f=4.15$ MeV, the pre-equilibrium fission, which anyway can not reproduce the fore-aft asymmetry, is insignificant.

With a decrease of $E_{cm}$ from 75.6 MeV to 72.8 MeV $(A_{exp}-1)$ increases from $\approx 0.85$ to $\approx 1.6$ \cite{Hinde96}. In contrast,
since $<J^2(75.6 {\rm MeV})>/<J^2(72.8 {\rm MeV})>\approx 1.6$,
the TSSM with $K_0^2\approx 200$ predicts a decrease of $(A_{TSSM}-1)$ from $\approx 0.16$ to $\approx 0.1$ \cite{Hinde96}. We take $<J^2(72.8 {\rm MeV})>=80$ and $<J^2(75.6 {\rm MeV})>=128$.
Then, to reproduce $A_{exp}$ within the TSSM one has to take $K_0^2(75.6 {\rm MeV})\approx 37$ and
$K_0^2(72.8 {\rm MeV})\approx 12$ instead of its constant value $K_0^2\approx 200$ with ${\cal J}_{eff}$ calculated within the model \cite{Sierk86}.

With  $T_a^J\propto \exp[-J^2/(<J^2>]$, $K_0^2=\infty$, $\Phi=\pi$ the mass-summed FFAD, obtained by symmetrizing Eq.~\eqref{Wcorrect} about $\theta=\pi/2$, depends on ${\dot \Phi}$ and $\beta/\Gamma$. To minimize a width of $P_{K=0}(\theta )$ for the deeply sub-barrier energy $E_{c.m.}=72.8$ MeV we take
${\dot \Phi}=0$ resulting in $\Delta_{K=0}\approx 6^\circ$. The fit of $A_{exp}(72.8 {\rm MeV}) =2.6$ in Fig. 1 uniquely yields $\beta/\Gamma=1.66$. For this same value of $\beta/\Gamma=1.66$
$A_{exp}(75.6 {\rm MeV}) =1.85$   is reproduced in Fig. 1 with
$|{\dot \Phi}|=3.6^\circ $ corresponding to a width of $P_{K=0}(\theta )$ of $\approx 8^\circ$ with its tail extended up to $\approx 135^\circ$.
This demonstrates that a fast growing of $A$ with the energy decrease
is due to the moderate reduction
of a dispersion of the pre-fusion dinuclear orientation with a decrease of $E_{c.m.}$.
For $T_b=1$ MeV and $B_f=4.15$ MeV, $\Gamma\approx 2.5$ keV. Then, for $\beta/\Gamma =1.66$, we find
$\beta\approx 4$ keV$\ll\Gamma_{spr}\simeq 5$ MeV.

% \begin{figure}[htbp]
  \begin{figure}[ht]
    \includegraphics[height=5.3cm]{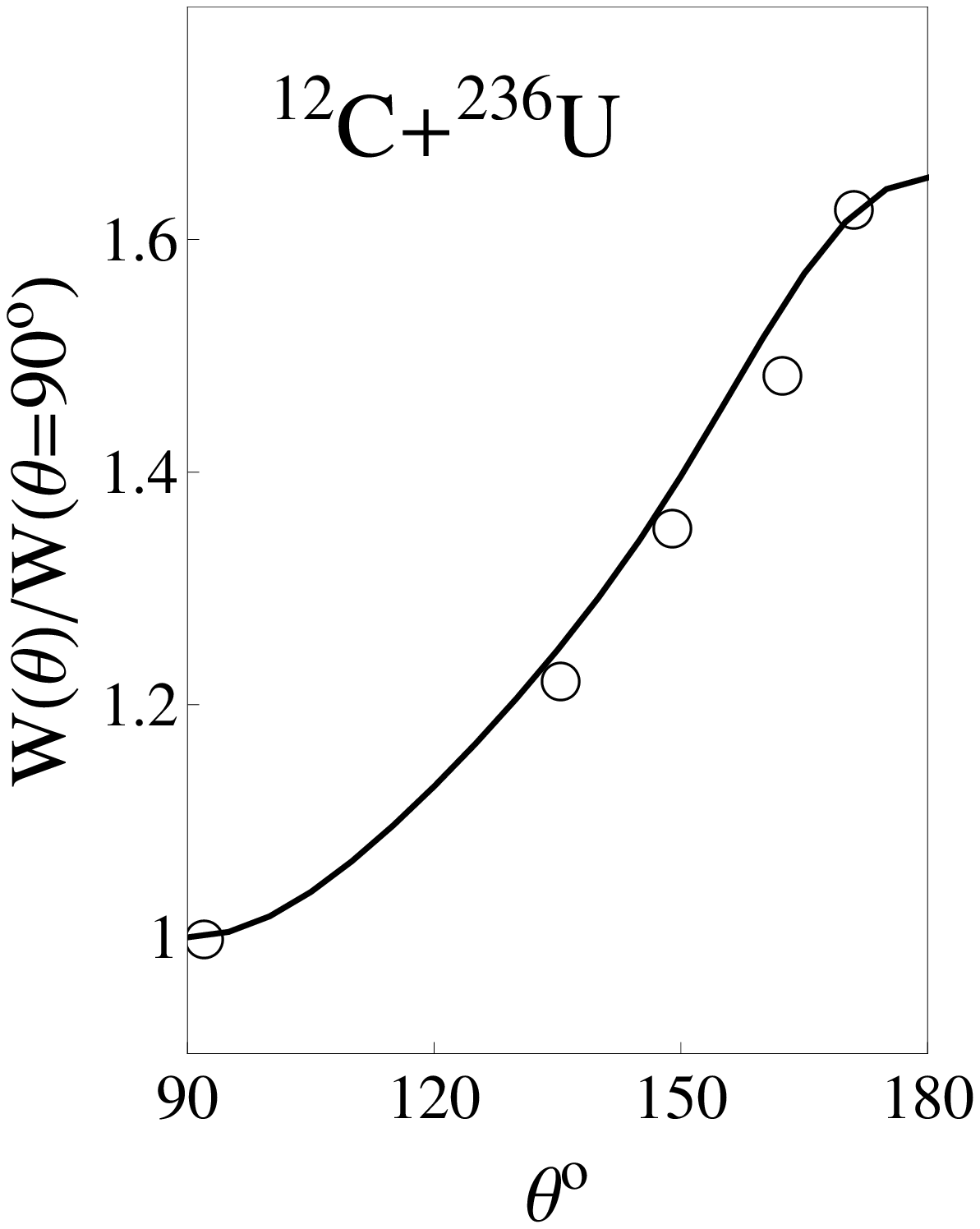}
    \includegraphics[height=5.3cm]{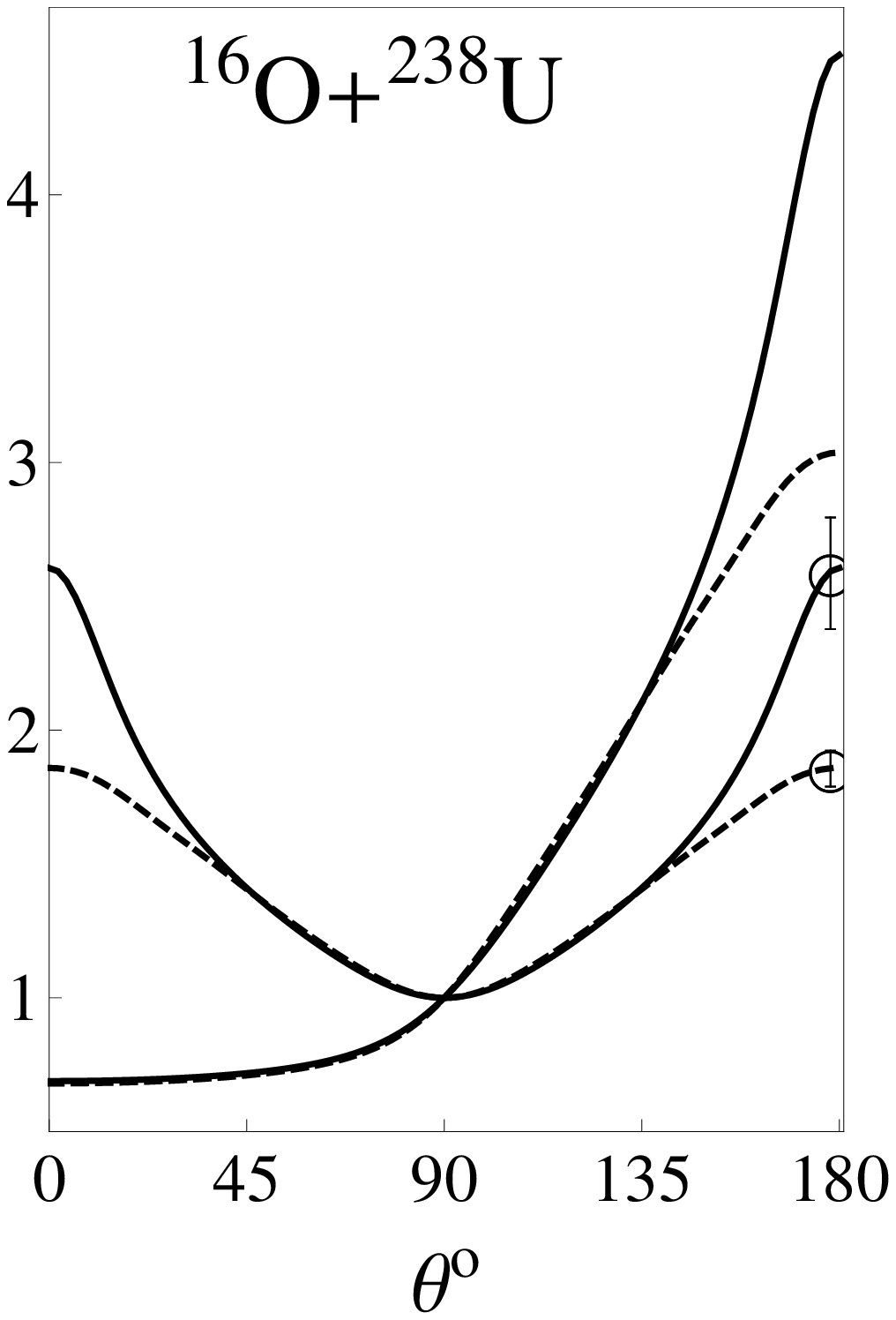}
 %   \centerline{ a) \hspace{5 cm} b)}
	\caption{Left panel: Mass-summed FFAD for $^{12}$C+$^{236}$U collision with $E_{c.m.}=59$ MeV. The data are from
\cite{Vandenbosch86}.
 The fit is obtained with $K_0^2=\infty$, $\beta/\Gamma=2.8$, $\Phi=\pi$, $|{\dot \Phi}|=3^\circ$ (see text).
Right panel: Solid and dashed curves are the calculated with $K_0^2=\infty$ FFAD for $^{16}$O+$^{238}$U collision with
 $E_{c.m.}=72.8$ MeV and 75.6 MeV, respectively. Symmetric and asymmetric about 90$^\circ$ curves correspond to FFAD
 for the light fragment masses and summed over the fragment masses, respectively (see text). Dots with error bars are the experimental
 anisotropies obtained from mass-summed FFAD \cite{Hinde96}.}
	\label{fig_1}
\end{figure}

In Fig. 1 we also display the unsymmetrized around $\pi/2$ FFAD's calculated for $\Phi=\pi$, $\beta/\Gamma=1.66$
($\beta=\beta_{\pi_1=\pi_2}= \beta_{\pi_1\neq \pi_2}$)   with ${\dot \Phi}=0$ ($E_{c.m.}=72.8$ MeV) and
$|{\dot \Phi}|=3.6^\circ $ ($E_{c.m.}=75.6$ MeV)
for the lighter fragments,
$M_L\approx 0.4(M_L+M_H)$, where $M_{L(H)}$ are masses of the lighter (heavier) fragments.
FFAD's for the heavier fragments, $M_H\approx 0.6(M_L+M_H)$,  can be obtained by the reflection around $\theta=\pi/2$, $W_H(\theta)=W_L(\pi -\theta )$, demonstrating the significant
mass-angle correlation observed in \cite{Hinde96}. Keeping $\beta_{\pi_1=\pi_2}=\beta=$const (to keep the mass-summed FFAD unchanged)
but increasing $\beta_{\pi_1\neq \pi_2}$ would result in a reduction of the fore-aft asymmetry in Fig. 1 and weakening of the mass-angle correlation.
For $\beta_{\pi_1\neq \pi_2}/\Gamma\gg 1$ the mass-angle correlation is destroyed resulting in the fore-aft symmetry of the FFAD's independent
of the fragment mass.

Suppose that, on the pre-fusion stage, the prolate shape of the target
nucleus becomes unstable changing to the oblate shape with the symmetry axis oriented either along or perpendicular to the beam direction.
In fact, one may not exclude a coexistence - a coherent superposition - of the three pre-fusion alternative dinuclear configurations contributing
into a formation of the same CN. It would be interesting to see if interference between the three corresponding fusion-fission amplitudes would produce
oscillations in the FFAD.

In case of the noticeable suppression of the evaporation residues yield indicating
quasifission   one would be led to deal with the  $(J_1\neq J_2 )$-correlation
at the conditional saddle-point. The corresponding expression for the angular distribution can be envisaged from the
consideration of the angular distributions in dissipative heavy-ion collisions \cite{Kun97b}. The result for the
time power spectrum is given by Eq.~\eqref{Wcorrect} but with
$\exp(-\Gamma t/\hbar)\exp[-i\omega t(J_1-J_2)-\beta |J_1-J_2|t/\hbar]$ instead of $[1+(\beta /\Gamma )|J_1-J_2|]^{-1}$,
where $\omega$ is a real part of the angular velocity of the coherent rotation of the dinuclear system. In this expression
one can easily take into account a possible time-dependence of ${\cal J}_{eff}$ (and, therefore, $\omega$) and $K_0^2$.
For $\beta \ll\Gamma$ and $\Phi\approx 0$ the resulting angular distribution
is $\propto\cosh [(\pi -\theta)\Gamma /(\hbar\omega )] $ demonstrating that the quantum-mechanical
$(J_1\neq J_2 )$-interference produces a classical-mechanics picture of
the rotating osculating complex with a classically single angular momentum,
Fig. 14 in \cite{Herschbach86}.

The wave nature of the heated organic macromolecules has been firmly established, in a model-independent way, raising a question of matter-wave interference
for biologically functioning entities of elevated temperature carrying the code of self-replication such as viruses and bacteria \cite{Geyer16}.
In this Letter we have proposed that the essentially wave phenomenon - the CN spin off-diagonal correlation -
plays an important role in nuclear fission.
We have demonstrated that
the mass-angle correlation does not necessarily signify hindrance of the CN formation due to
the quasifission but can originate from the CN spin off-diagonal phase correlation in the fusion-fission.
One of the central conventionally counter-intuitive problems of the proposed interpretation, which has not been explicitly addressed in the study of matter-wave interference with complex molecules \cite{Geyer16}, is to justify
the anomalously slow cross-symmetry phase relaxation
in classically chaotic many-body systems as compared to the relatively fast phase and energy relaxation (thermalization) within a single symmetry sector.
Seemingly paradoxically on the first sight, this problem is intimately related \cite{KunInProgress} to
the long-standing Wigner dream \cite{Gard72} to develop a theory of correlations between reduced widths within
a single symmetry sector $(J_1=J_2,\pi_1=\pi_2)$. The subject also relates to the problem of the quantum-classical transition in macro-world.

\begin{acknowledgments}

\end{acknowledgments}

% Create the reference section using BibTeX:
%\bibliography{basename of .bib file}
%%\begin{thebibliography}{apssamp.bib}

\end{document}